# Unraveling the Complexity of Metal Ion Dissolution: Insights from Hybrid First-Principles/Continuum Calculations

Mingqing Liu[1], Tong-Yi Zhang[1,2*], Sheng Sun [1,2,3*]

1. Materials Genome Institute, Shanghai University, Shanghai 200444，China

2. Shanghai Frontier Science Center of Mechanoinformatics, Shanghai University, Shanghai 200444, China

3. Zhejiang Laboratory, Hangzhou 311100, China

Corresponding authors：mgissh@t.shu.edu.cn(S.S)，zhangty@shu.edu.cn(T.Y.Z.)

**Abstract**：The study of ion dissolution from metal surfaces has a long-standing history, wherein the gradual dissolution of solute atoms with increasing electrode potential, leading to their existence as ions in the electrolyte with integer charges, is well-known. However, our present work reveals a more intricate and nuanced physical perspective based on comprehensive first-principles/continuum calculations. We investigate the dissolution and deposition processes of 22 metal elements across a range of applied electrode potentials, unveiling diverse dissolution models. By analyzing the energy profiles and valence states of solute atoms as a function of the distance between the solute atom and metal surface, we identify three distinct dissolution models for different metals. Firstly, solute atoms exhibit an integer valence state following an integer-valence jump, aligning with classical understandings. Secondly, solute atoms attain an eventual integer valence, yet their valence state increases in a non-integer manner during dissolution. Lastly, we observe solute atoms exhibiting a non-integer valence state, challenging classical understandings. Furthermore, we propose a theoretical criterion for determining the selection of ion valence during electrode dissolution under applied potential. These findings not only contribute to a deeper understanding of the dissolution process but also offer valuable insights into the complex dynamics governing metal ion dissolution at the atomic level. Such knowledge has the potential to advance the design of more efficient electrochemical systems and open new avenues for controlling dissolution processes in various applications.

**Keywords**：Metal dissolution/deposition, valence state, charge transfer

## 1 Introduction

The notion of ions was first introduced by Faraday, who postulated that the application of an electric current to an electrolyte solution would induce the motion of discrete entities known as ions. Building upon Faraday's work, Arrhenius put forth the theory of ionization, which proposed that ions inherently existed within electrolyte solutions, even in the absence of external electric potentials. According to Arrhenius, the role of an external electric potential was to impart directionality to ion movement rather than create new ions. This theory formed the bedrock of electrochemistry and garnered widespread acceptance. Arrhenius postulated the existence of two types of ions in electrolyte solutions, each carrying equal but opposite charges, a view widely supported by scholars. It is conventionally assumed that ions solely exist in integer valence states in the electrolyte solution. To date, no experimental or theoretical evidence has definitively confirmed whether ions in electrolyte solutions exclusively adopt specific integer valence states, as upheld by conventional wisdom, or if they may manifest in alternative forms.

Recent research has unveiled intriguing possibilities regarding the valence states of atoms when adsorbed on electrode surfaces[1]. It has been observed that these valence states may deviate from traditional integer values. Rather, the adsorbed atom can form polar or covalent bonds with the electrode, sharing electrons and yielding a non-integer electrical valence, known as the adsorption valence. Marichev et al.[2] made a pioneering observation through in situ contact resistance technology, enabling the measurement of the partial charge of surface-active anions at the solid-liquid interface. Contact resistance was measured by periodic contacting two samples under a constant potential in a fixed electrolyte. The results conclusively demonstrated that the presence of surface-active anions on silver metal electrodes led to an increment in charge transfer with a positive potential shift until complete charge transfer occurred from the anion to the electrode. The potential corresponding to complete charge transfer depended on various factors, such as the metal type, concentration, solvent, and anion. A comprehensive review by Schmickler and Guidelli[1] further postulated that partial charge transfer may arise from electron flow between metal surface atoms and adsorbed molecules, facilitated by the overlapping of metal orbitals and atomic orbitals within the adsorbed molecules. In cases where a complete ionic bond forms between the metal and adsorbed molecules, the partial charge transfer coefficient becomes zero, precluding electron flow between them. Notably, Grunder et al[3]. employed resonant surface X-ray diffraction to directly detect partial charge transfer between metal electrodes and adsorbed species in electrolyte solutions. By coupling the polarization of the incident X-ray beam with the electron density at the interface, they discerned significant alterations in charge distribution for both surface and subsurface metal atomic layers during the formation of ionic bonds, as evidenced by the halide layer on single crystal Cu(001) and Au(001) electrodes.

    The valence state of ions in electrolyte solutions was still a challenge topic and yet unclear. The chemical method is commonly used to analyze ion valence states. It has been developed with direct potentiometry into potentiometric titration, a widely used technique. In potentiometric titration, an oxidizing or reducing agent is added, and changes in potential determine the redox potential and qualitatively analyze ion valence states. However, it is limited to qualitative analysis and lacks quantitative information. The gravimetric method[4] is used for

quantitative analysis in electrochemistry. The substance is weighed and subjected to electrolysis in the electrolyte. Mass changes are measured, and integrating the current yields the charge. Dividing charge by mass difference provides the average ion valence state. Drawbacks include imprecise measurements and uncertainty in ion states. Determining ion valence states in electrolytes remains challenging due to difficulties in capturing ions and determining valence without altering net charge distribution. Precise measurement is almost unattainable.

The electro-dissolution and electro-deposition of ions at the interface between solids and liquids constitute a fundamental process and serve as an ideal model for investigating variations in ionic valence. While the dynamic changes in the reaction mechanism at the solid-liquid interface during these processes have garnered significant attention for their profound implications, the intricate reaction environment and the intricate interplay between electrons and ions pose considerable challenges for investigation. Traditional experimental approaches at the atomic scale encounter limitations in characterization and observation, making computational methods utilizing density functional theory (DFT) an attractive alternative.

For instance, Sharma et al.[5] proposed a three-step model for the dissolution of solute atoms from a metal surface into the electrolyte: the breaking of metal bonds between adsorbed atoms and surface atoms, the formation of an ionic solvation shell, and the migration of solute atoms in the electric double layer. Through their study, they observed a gradual change in the valence state of the $Al^{3+}$ hexahydrate complex from +2 to +3, while the valence state of the Al ion at the center of the complex remained stable around +2.6. This investigation revealed that the variation in ion valence primarily manifests as charge transfer within the surrounding adsorbed water complex.

While computational methods offer advantages in circumventing experimental challenges, the electrochemical processes occurring at the solid-liquid interface remain a complex research problem encompassing aspects such as the solid-liquid interface, space charge, and surface structure. Atomic-scale DFT calculations, although powerful, fall short in fully capturing the electrolyte environment, while classical molecular dynamics lack effective means to simulate high-precision electron transfer. Moreover, macroscopic continuum models fail to characterize the local characteristics of individual atoms. Consequently, the development of joint density functional theory[6] has emerged as the most promising numerical method for investigating the solid-liquid interface, particularly when charge is applied. This approach combines quantum DFT for the electrode with a continuum model incorporating non-local dielectric response for the solution, enabling accurate simulations of the electrochemical processes at the solid-liquid interface. For example, Jiang et al.[7] employed joint density functional theory to investigate the phenomenon of partial charge transfer during the electro-dissolution and electro-deposition of Zn atoms on the Zn (0001) crystal face electrode. Their study exemplifies the power of this combined approach in elucidating the intricate dynamics of charge transfer at the solid-liquid interface.

In this paper, we embark on a comprehensive exploration of ionic valence variations during electro-dissolution and electro-deposition at the solid-liquid interface. Leveraging joint density functional theory, we identifies three distinct dissolution models for different metals and challenging the long-held assumptions. Our findings have the potential to reshape our fundamental comprehension of ions in electrolyte solutions and pave the way for exciting advancements in the field of electrochemistry.

## 2 Computational methods

The calculations were performed using JDFTx[6], an open-source computing software based on plane wave basis expansion. JDFTx[6] couples the DFT calculation of the electrode with adsorbed atoms and the continuum solution models through spacial distributed electropotential. To maintain the electrical neutrality of the primitive cell in simulations with periodic boundary conditions (PBC), the applied charge of explicit atoms and the total induced binding charge in the continuum electrolyte are balanced. It's worth noting that the continuum model in JDFTx[6] does not account for the reactivity of the solution, providing a chance to apply virtual experiments to identify charge transfer solely between electrode and dissolved ions.

The calculation model is depicted in Figure 1, which illustrates the dissolution of solute atoms from the stable adsorption point on the electrode surface into the electrolyte. The computing system comprises a 5-layers thin plate electrode of 20 atoms with each layer consisting of 4 atoms and 2 adsorbed atoms on both sides, forming a plane-symmetric structure.

The supercell model and PBC were employed for the calculations, with both sides of the thin plate electrode filled with electrolyte. The electrolyte implemented the SaLSA (The Spherically Averaged Liquid Susceptibility Ansatz) continuum model[8], with a concentration of 1M $Na^+$ and $F^-$. The exchange correlation function was chosen to be the generalized gradient approximation[9] (GGA), while the functional was set to PBEsol[9]. The GBRV ultrasoft pseudopotential[10] was utilized. Here, we only present the computed data for 8 representative metal elements. The data for other metals can be found in the supporting information. The K point settings for the electrodes of K (100), Mg (0001), Al(111), Ge (111), Y(0001), Cu(100), Ag(100) and Sc(0001), are 6×6×1, 12×12×1, 10×10×1, 4×4×1, 1×1×1, 4×4×1, 12×12×1, 12×12×1 and 4×4×1, respectively, with a cut-off energy of 598 eV. The relative accuracies in the iteration energy differences of electron cloud density, atomic (ionic) coordinates, and lattice constant were set to be $2.72 \times 10^{-7}$, $2.72 \times 10^{-7}$, and $2.72 \times 10^{-8}$ eV, respectively.

Structural optimization is performed on the buildup systems. Firstly, the thin plate electrode model without solute atoms is relaxed successively in both vacuum and the electrolyte environment. During this process, the lattice size and atomic position are optimized, and the energy of the structure is minimized. Next, adsorbed atoms are added to the electrode surface, as shown in Figure 1, and its energy is minimized through free relaxation. Hereinafter, the added adsorbed atom is called the solute atom. This step ensures that the solute atom can reach the stable adsorption point. Subsequently, all the atoms on the electrode of the thin plate are fixed, and the distance between the solute atoms and the electrode surface is manually adjusted under different applied charge levels.

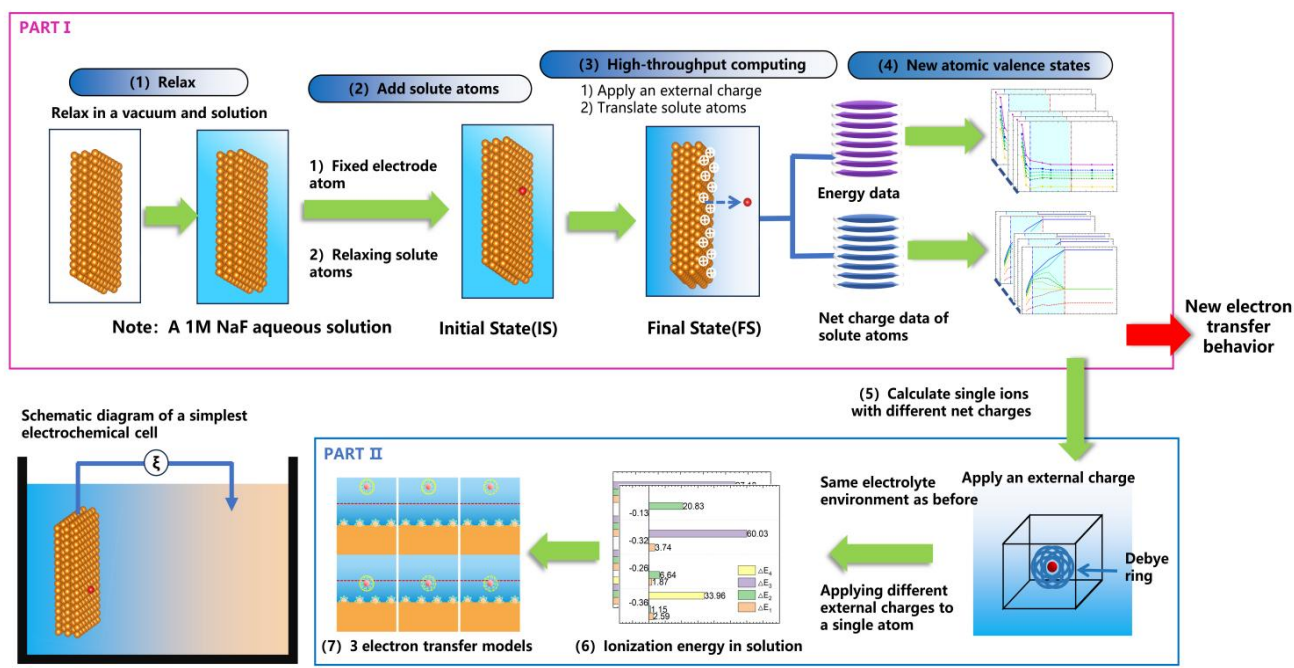

**Figure 1**: Schematic diagram illustrating the research workflow and simulation processes of dissolution/deposition of adsorbed atoms. The diagram provides an overview of the steps involved in the research process and demonstrates the simulation procedures for the dissolution and deposition of adsorbed atoms.

### 3 Results and discussion

By carefully manipulating the spacing between solute atoms and the electrode surface, as well as adjusting the total charges of both the electrode and solute-atom system, we simulated the dissolution dynamics of different metal elements within an electrolytic environment as a response to applied charges. Through an exhaustive assessment of energy profiles and partial charges associated with solute atoms during the dissolution process, utilizing Löwdin population analysis, we have delineated three distinctive classifications that elucidate the dissolution characteristics of metal ions. This discovery sheds light on an array of phenomena that extend beyond the confines of our conventional comprehension.

**3.1 Category #1: The dissolved valence state aligns with the conventional understanding, exhibiting an integer jump in the valence state.**

Figure 2 illustrates the energy profiles of dissolution/deposition (a, c, and e) and their corresponding partial charges (b, d, and f), contingent upon the distance between solute atoms and electrode surfaces. Our analysis centered on solute atoms of potassium (K) and magnesium (Mg) on the K(100) and Mg(0001) surfaces, respectively, with variable applied charges. Within the energy profiles, a solid line signifies the absence of an energy barrier along the reaction path, indicative of spontaneous dissolution. Conversely, a dashed line denotes a stable adsorption state. For Mg, the position of stable adsorption, if present, was identified at $z = 0.856$ Å, with minimal variance noted under differing applied charges. This location was referenced with respect to the outermost surface layer established as $z = 0$. The light blue segment in Figure 2 designates the transitional realm for energy and partial charge. When z surpasses the right boundary of this transition area, neither energy nor partial charge exhibits further variation. This demarcated area being right of the transition region is termed the energy-stable region, implying a complete dissolution of ions into the electrolyte.

The solute K atom situated on the K metal electrode undergoes spontaneous dissolution into the electrolyte, displaying no encounter with an energy barrier, even under an applied charge of $0.0|e|$, as visualized in Figure 2(a). Upon dissolution, the solute K atom acquires an integer valence of +1 within the energy-stable region of the electrolyte, as illustrated in Figure 2(b), corroborating our conventional understanding. Additionally, we noted a direct shift of the solute K atom's valence from the adsorption charge to +1, devoid of any non-integer partial charge in the energy-stable electrolyte region. Our computations also reveal that elements akin to K, such as lithium (Li) and sodium (Na), demonstrate analogous behavior (refer to Figure S1(a,b) and S2(a,b) in the Supplementary Materials).

In contrast to K, Mg ions exhibit a lack of spontaneous dissolution behavior when the applied charge remains at or below 1.0|e|/22 atoms, as illustrated in Figure 2(c). These findings imply that the reactivity of Mg is less pronounced compared to that of K, aligning with our conventional understanding. As the applied charges surpass 2|e|/22 atoms, the solute Mg atom naturally dissolves into the electrolyte, embracing an integer valence of +2 within the energy-stable region (Figure 2(d)), congruent with both experimental evidence and established knowledge. This underscores the capacity of first-principles calculations to uncover the underlying reaction mechanisms within electrochemical dissolution/deposition processes. Our calculations further unveil that elements akin to Mg, such as beryllium (Be) and calcium (Ca), manifest analogous behaviors (refer to Figure S3(a,b) and S4(a,b) in the Supplementary Materials).

Figure 2(e, f) portrays the dissolution behavior of the solute Al atom, which becomes spontaneous once the applied charge surpasses 4.5|e|/22 atoms. Within the energy-stable region of the electrolyte, the solute Al atom adopts an integer valence of +3. However, in cases where applied charges are lower than 4.5|e|/22 atoms, the solute Al atom fails to dissolve autonomously due to the presence of an energy barrier. The dissolved valence value aligns seamlessly with our classical comprehension. Significantly, when manually introduced into the electrolyte, the solute Al atom consistently exhibits a valence state of +1. Though this +1 valence state lacks physicality within the energy-stable region, the findings propose a distinctive dissolution mechanism for Al when contrasted with K and Mg. Moreover, the results suggest the potential occurrence of Al with a +1 valence state under specific conditions, such as varying electrolytes or electrode surfaces, warranting further exploration.

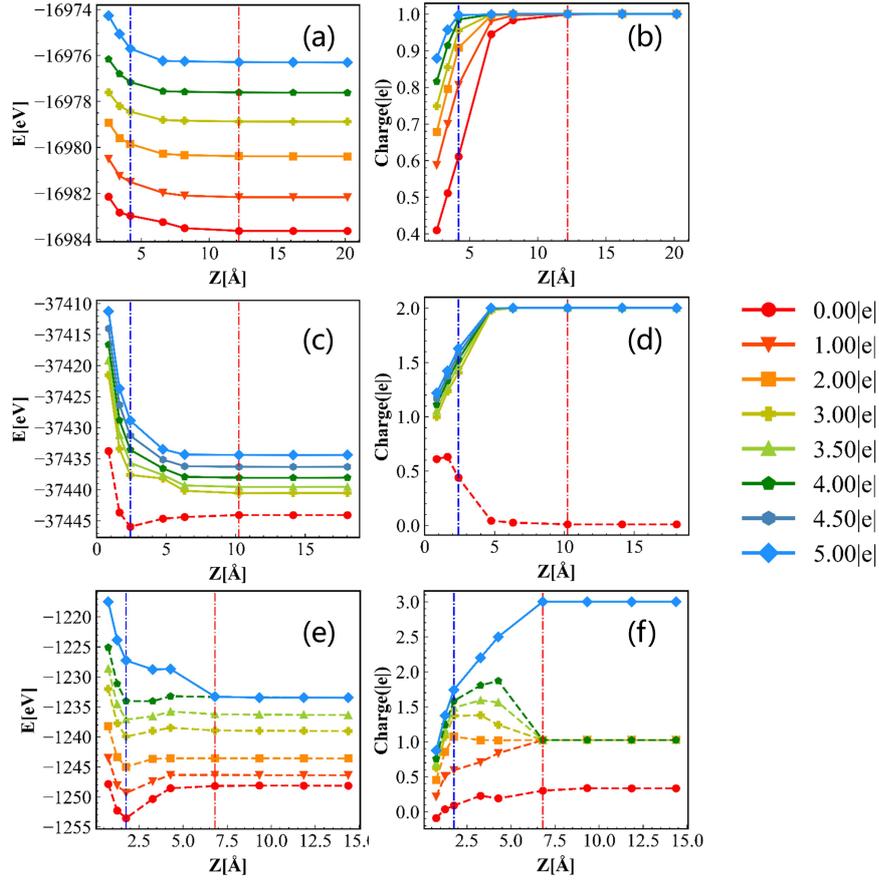

**Figure 2:** Energy profiles and changes in valence states of K, Mg, and Al electrodes during the dissolution/deposition process.

**3.2 Category #2: The dissolved valence state adheres to the conventional understanding, while the valence variation does not demonstrate a straightforward integer transition.**

Figure 3 illustrates the energy profiles and the profiles of solute atomic valence during the dissolution/deposition process of Y and Ge on the Y(0001) and Ge(111) surfaces, respectively. In contrast to the three metals discussed earlier, where the valence state changes by an integer during dissolution, involving the loss of one electron per solute atom, the Y and Ge electrodes deviate from this pattern.

Regarding the Y electrode, when the applied charge equals or surpasses 4.0|e|/22 atoms, the

energy barrier along the reaction path diminishes. The solute Y atom spontaneously dissolves into the electrolyte, adopting an integer valence state of +3. These findings are in alignment with conventional understanding. Conversely, with applied charges less than 4.0|e|/22 atoms, an energy barrier emerges in the reaction path. In contrast to Al ions, which converge to a +1 valence state during the virtual dissolution process, Y ions demonstrate a non-integer valence state. Under applied charges of 0.00, 1.00, and 2.00 |e|/22 atoms, the valence states are -0.069, +0.467, and +0.666 |e|, respectively, in the energy-stable region. Interestingly, Y displays non-integer charges solely when the net charge remains below +1, transitioning to an integer within the range of [+1, +3]. This observation suggests a mixed charge transfer behavior, possibly indicating distinct charge transfer mechanisms at play.

When the applied charge exceeds 5|e|/22 atoms, the solute Ge atom dissolves spontaneously with a valence state of 2.0|e| in the energy-stable region. These findings align with established knowledge. Energy barriers emerge on the corresponding reaction path when the applied charge is below 5|e|/22 atoms. Intriguingly, the valence state of the solute Ge atom exhibits non-integer variations. Specifically, the valence states are +0.125, +0.565, +1.136, and +1.736 |e| under applied charges of 0.00, 1.00, 2.00, and 3.00 |e|/22 atoms, respectively, in the energy-stable region. This observation challenges the conventional notion of valence states undergoing integer transitions for isolated atoms. Our calculations also disclose that elements akin to Ge, such as Zn, Cd, and Pb, exhibit analogous behavior (refer to Figure S9 - S13 in the Supplementary Materials).

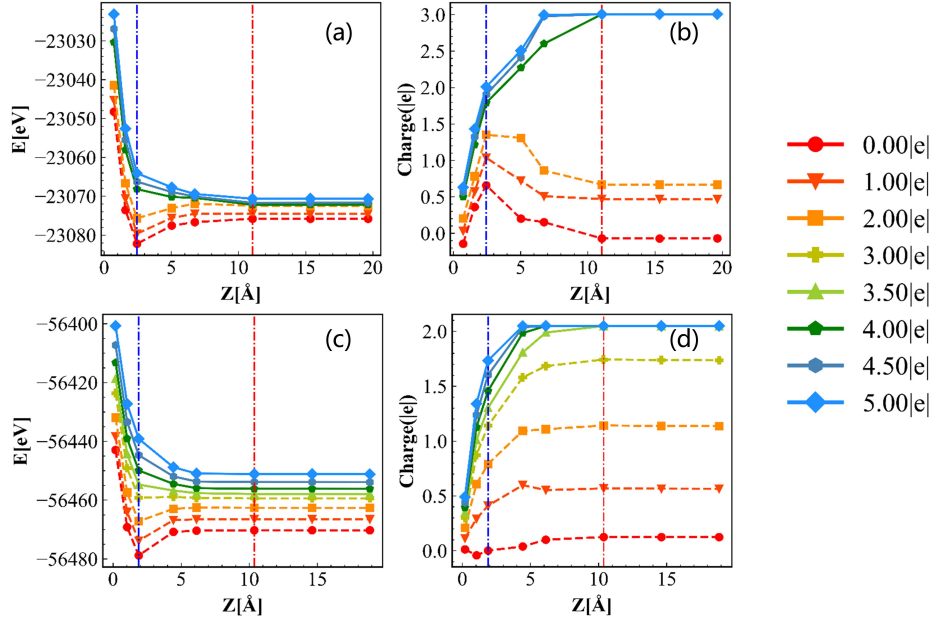

**Figure 3**: Energy profiles and changes in valence states of Ge and Y electrode during the dissolution/deposition process.

**3.3 Category #3: Charge-dependent non-integer valence states after dissolution, challenging the conventional understanding.**

Figure 4 illustrates the energy profiles and valence profiles of solute atoms during the dissolution/deposition process of Cu and Ag from/to Cu(100) and Ag(100) surfaces. In contrast to the five elements mentioned previously, the valence states of the solute atoms persist as non-integer charges even upon dissolution into the electrolyte.

As the applied charges exceed 3.5|e|/22 atoms, the solute Ag atom spontaneously dissolves into the electrolyte, evident by the monotonically decreasing energy profiles. Conversely, when the applied charges fall below 3.5|e|/22 atoms, an energy barrier emerges on the corresponding reaction pathway, leading to a non-integer shift in the valence state of the solute Ag atom. This behavior mirrors the trends observed in the solute Ge atom. For applied charges of 3.5 and 4.0|e|/22 atoms, the solute Ag atom adopts an integer valence state of +1.0 in the energy-stable

region, in harmony with our conventional understanding. Intriguingly, beyond 4.0|e|/22 atoms, the valence state of Ag ions doesn't stabilize at +1, but rather continues to rise, resulting in non-integer charges. Specifically, the valence states in the energy-stable region under applied charges of 4.5|e| and 5.0|e|/22 atoms are +1.11 and +1.23|e|, respectively.

Concerning the Cu electrode, when the applied charge equals or exceeds 4.0|e|/22 atoms, there is an absence of an energy barrier on the reaction path. Intriguingly, the solute Cu atom undergoes spontaneous dissolution into the electrolyte while maintaining a non-integer valence state. The valence states within the energy-stable region, under applied charges of 4.00, 4.50, and 5.00|e|/22 atoms, are +1.416, +1.499, and +1.588|e|, respectively. It's perplexing that even in the absence of an energy barrier on the reaction path, Cu ions persist as non-integer charges. We conjecture that this may illuminate a charge transfer mechanism distinctly different from our traditional understanding. Furthermore, our computations indicate that elements akin to Cu, such as Sc, exhibit analogous behavior (refer to Figure 4(e,f)).

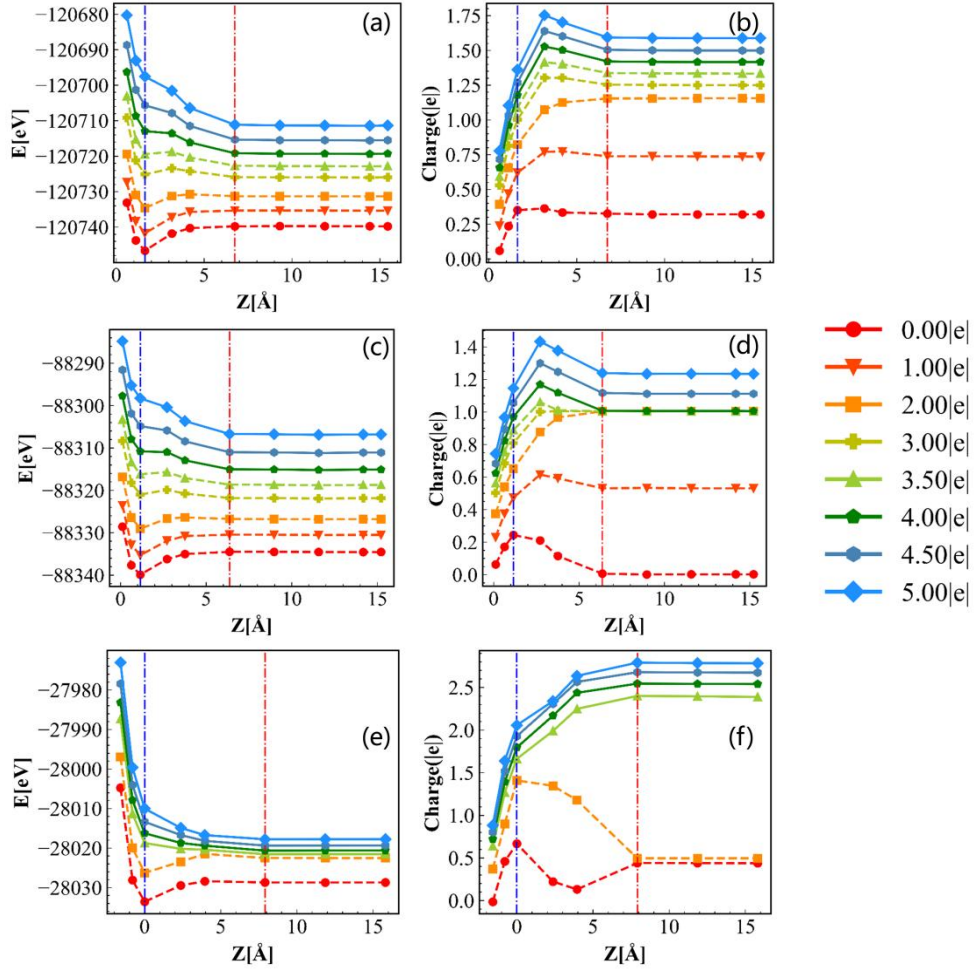

**Figure 4:** Energy profiles and variations in the valence states of (a, b)Cu, (c, d)Ag and (e, f)Sc electrodes during the dissolution/deposition process.

### 3.4 Further discussions

Given the utilization of PBC in all our calculations, a plausible question arises about whether the interactions between solute atoms in the primary simulation cell and their image cells contribute to the emergence of non-integer valence states. To address this, we conducted validation calculations on elements that had previously undergone first-principles calculations. We expanded the primary cell into a 3-layers thin plate electrode, encompassing 27 atoms, with each layer composed of 9 atoms and 2 adsorbed atoms on both sides. The results of these computations are presented in Table S1 in the Supplementary Materials, and the corresponding energy and valence state profiles are also detailed there.The outcomes indicated that the size of the supercell does not impact the fundamental findings. Specifically, we confirm the existence of three categories of metals regarding their valence state variations during dissolution. Minor quantitative deviations in the valence states of Zn and Cd were observed, but they do not undermine the overarching conclusions.

Utilizing the data acquired from our calculations, we have discerned distinct dissolution patterns among the dissolved metals, leading to the formulation of three distinct dissolution models. The initial category encompasses Li, Be, Na, Mg, Al, K, Ca, and Y, denoted by the orange color in Figure 5. Within the electrolyte, solute atoms of these metals consistently adopt an integer valence state. Their valence state transitions exhibit discrete, integer jumps, aligning seamlessly with conventional understanding. The second classification involves Ge, Y, Zn, Cd, and Pb, distinguishable by the green color in Figure 5. For these metals, solute atoms also maintain an integer valence state within the electrolyte. However, the valence state shifts showcase intriguing non-integer variations, signifying continuous transitions, a characteristic departure from the discrete transitions in the first category. The third category encompasses Cu, Ag, and Sc, depicted in the blue hue in Figure 5. Within the electrolyte, solute atoms from these metals adopt non-integer valence states. Unfortunately, despite our efforts to examine the metal elements indicated in purple in Figure 5, conclusive outcomes were not attained due to the failure of calculations to converge. This indicates the necessity for further comprehensive inquiries into these specific metals in the future.

**Figure5:** Metal elements labeled with three types of dissolution models and metal elements exhibiting non-converging valence states on the periodic table.

### 3.5 Possible explanations on the appearance of three dissolution valence models

Traditionally, the first ionization energy of an atom serves as a quantifier for the minimal energy needed for the atom to shed an electron in a gaseous environment. Likewise, the second ionization energy represents the minimum energy required to detach a second electron. These parameters hold substantial significance in the analysis of compounds and the exploration of valence states. In this current study, we broaden our exploration to encompass the ionization energy or ionization potential of an ion in the electrolyte phase. This endeavor is driven by the objective of offering plausible explanations for the emergence of the three dissolution valence models.

To accomplish this, we utilized first-principles calculations based on JDFT for 22 distinct elements within a non-reactive 1M NaF electrolyte. In each calculation, a singular atom was placed within the electrolyte, and its electrons were systematically removed in a stepwise, integer-jump fashion. Figure 6 visually presents the energy discrepancies for an individual solute atom across different valence states, depicted as $\Delta E_c$

$$\Delta E_c = E_c - E_{c-1},$$

where $E_c$ and $E_{c-1}$ represent the system energy when a solute atom is in a valence state of $c$ and $c-1$, respectively. Results show three types of $\Delta E_c$ variations.

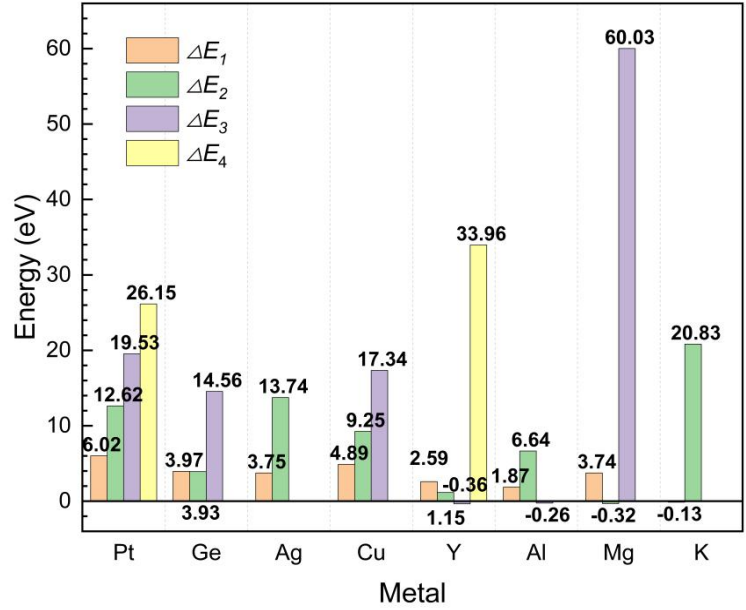

**Figure 6:** Comparative plot illustrating the different $\Delta E_c$ values for eight metal elements.

1) As shown in Figure 6, $\Delta E_1$, $\Delta E_2$ $\Delta E_3$, and $\Delta E_3$ are negative for K, Mg, Al and Y, respectively. The negative $\Delta E_c$ indicates that the corresponding $c$ th electron can be lost spontaneously in the electrolyte. We suggest that the spontaneous electron loss triggers an integer "jump" in the valence state of the solute atom. For instance, consider the cases of K, Mg, and Al solute atoms. These atoms exhibit a lack of non-integer valence states within the ranges of [0, +1], [0, +2], and [+1,+3], respectively, as illustrated in Figure 2.

2) The relative magnitude of $\Delta E_{c+1}$ over $\Delta E_c$ may indicate whether there exists an integer valence state. When $\frac{\Delta E_{c+1}}{\Delta E_c} > 3$, an integer valence state of $+c$ can be observed. For instance, the valence states of +1 for Al and Ag and +2 for Ge, whose $\frac{\Delta E_{c+1}}{\Delta E_c}$ are 3.55, 3.66 and 3.70, respectively. When $0 < \frac{\Delta E_{c+1}}{\Delta E_c} < 3$, non-integer valence states are always observed.

## 4 Conclusion

In this study, we identified three valence state models for metal electrode dissolution using hybrid DFT and continuum calculations for electrode and electrolyte, respectively. The first model aligns with classical understanding of integer valence state jumps. The second model involves integer dissolution but non-integer valence rise. The third challenges convention with non-integer dissolution states. We provide a theoretical criterion for predicting valence states based on ionization energies in electrolyte. Our results fill a theoretical gap in selecting atomic valence states during electrochemical dissolution/deposition, providing an important reference for studying charge transfer in these processes.

**Acknowledgements**

The work is supported by the National Natural Science Foundation of China (No. 12072179 and 11672168), and Key Research Project of Zhejiang Lab (No. 2021PE0AC02).

Supporting information：

**Category #1: The dissolved valence state aligns with the conventional understanding, exhibiting an integer jump in the valence state.**

Li(100)：

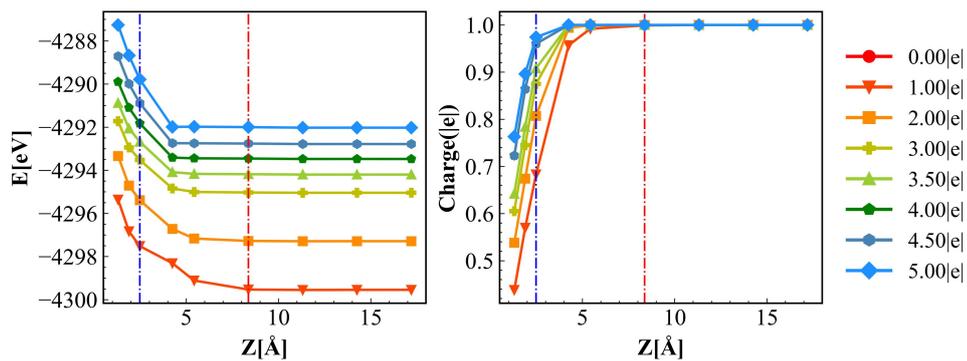

Na(100)：

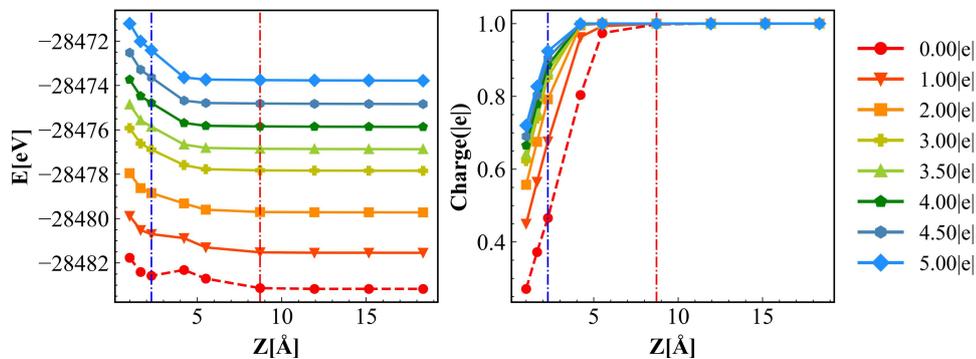

Be(0001)：

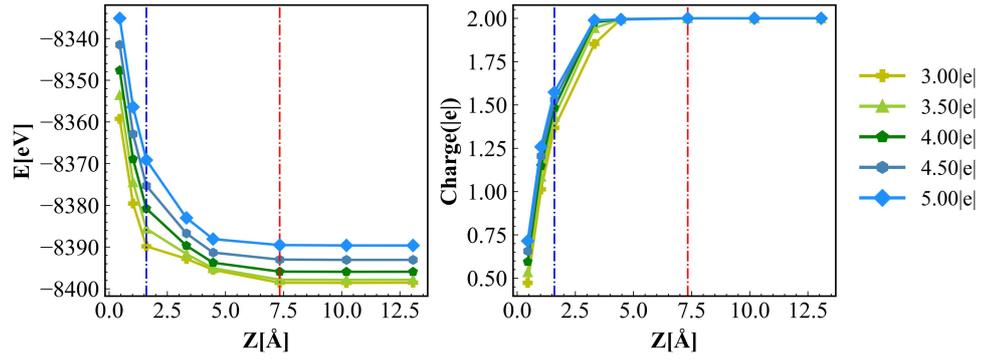

Ca(100):

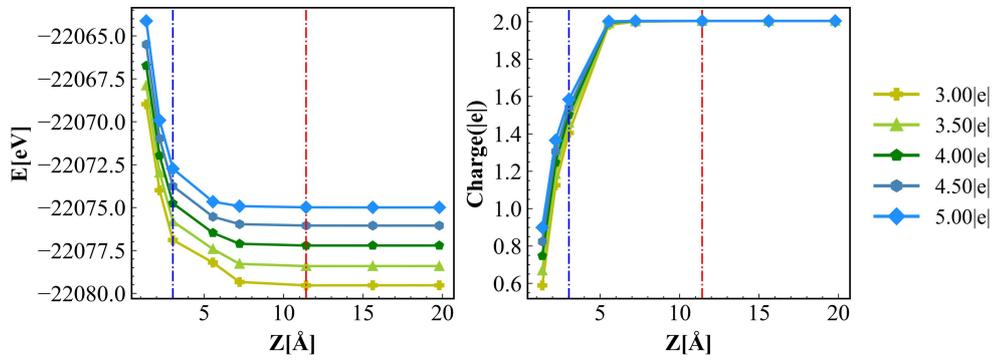

Mg(0001):

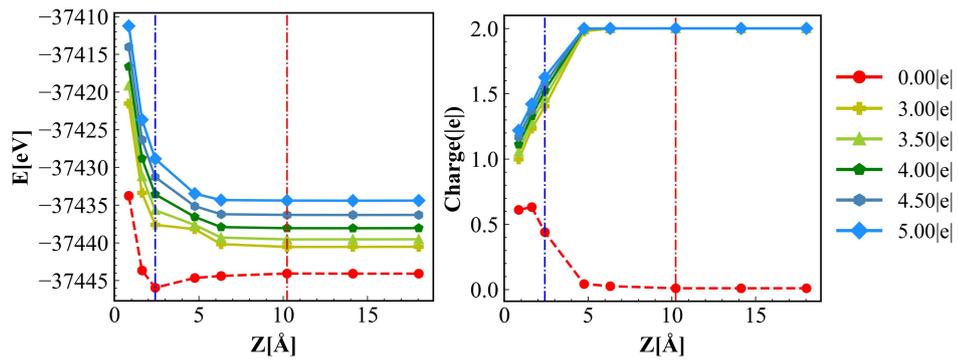

Enlarged supercell Mg(0001):

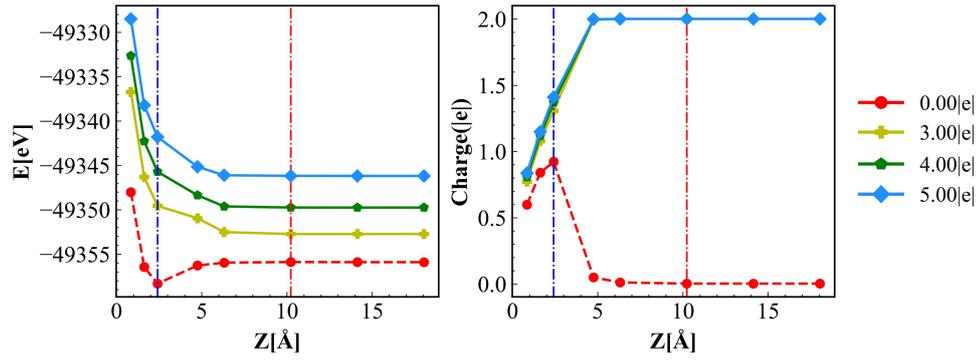

Al(100)：

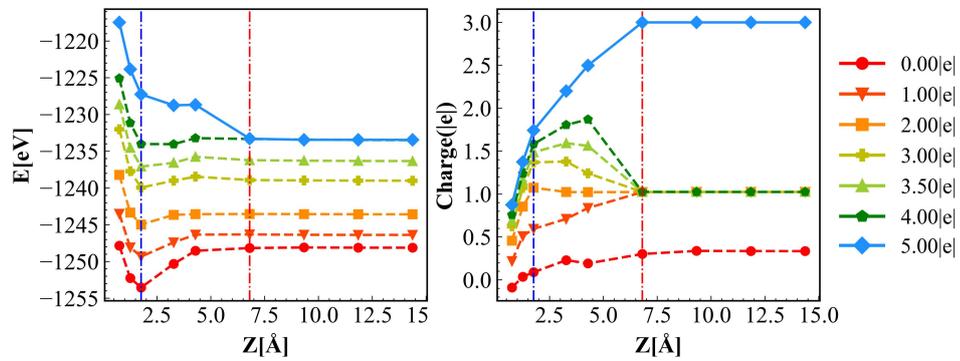

Enlarged supercell Al(100):

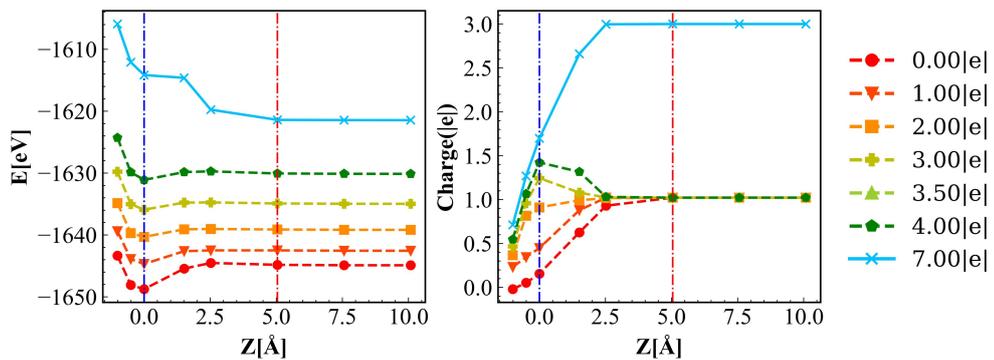

**Category #2: The dissolved valence state adheres to the conventional understanding, while the valence variation does not demonstrate a straightforward integer transition.**

Zn(0001):

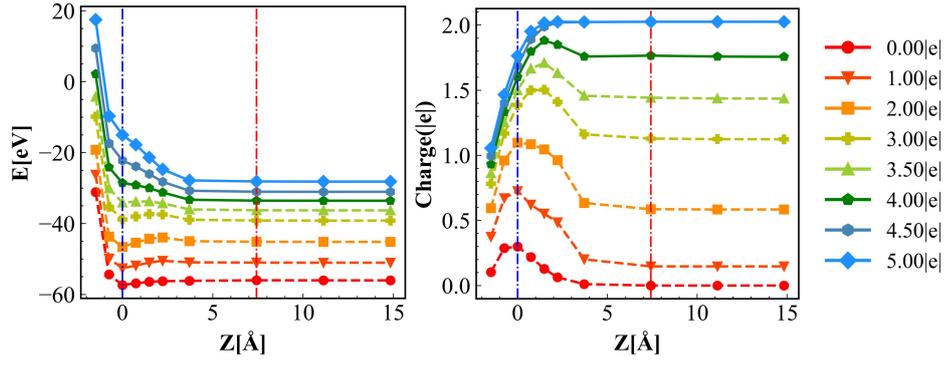

Enlarged supercell Zn(0001)：

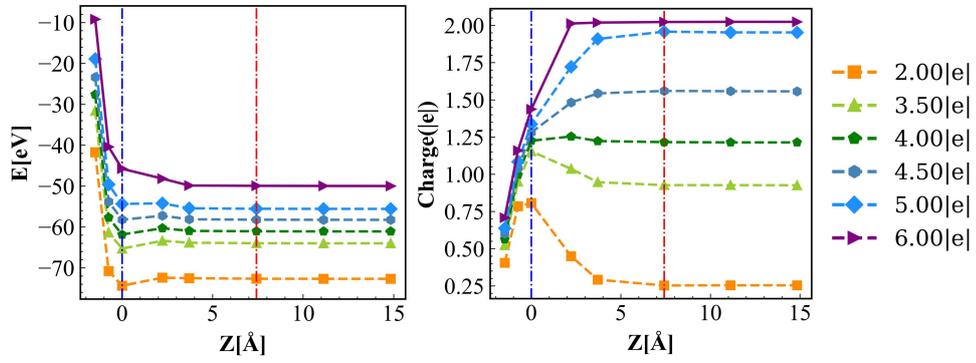

Cd(0001)：

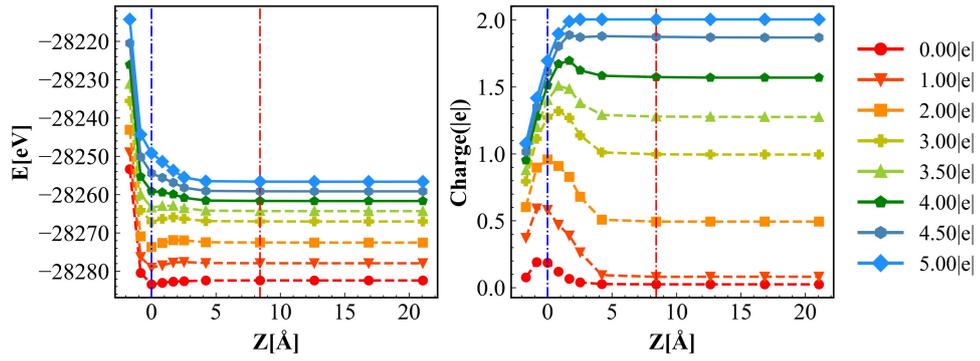

Enlarged supercell Cd(0001)：

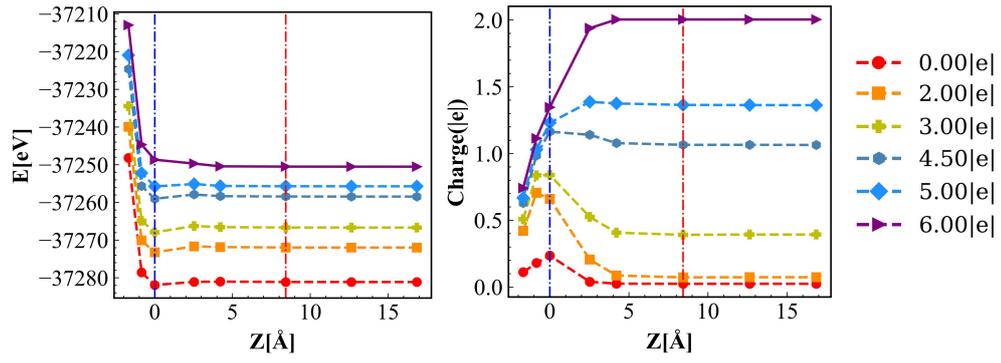

Pb(111):

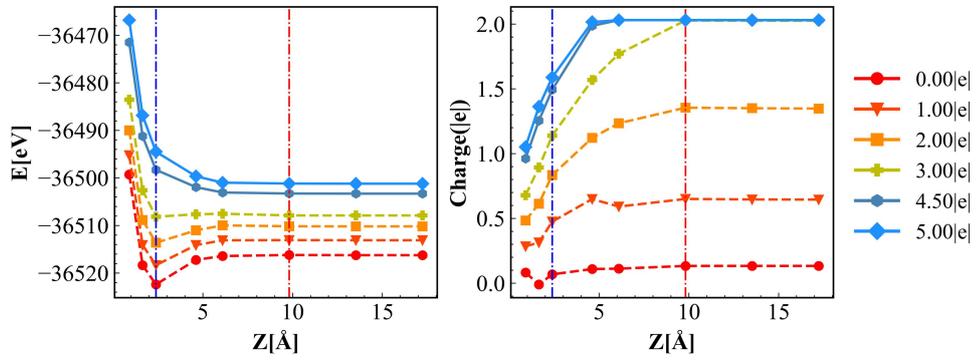

Ge(111):

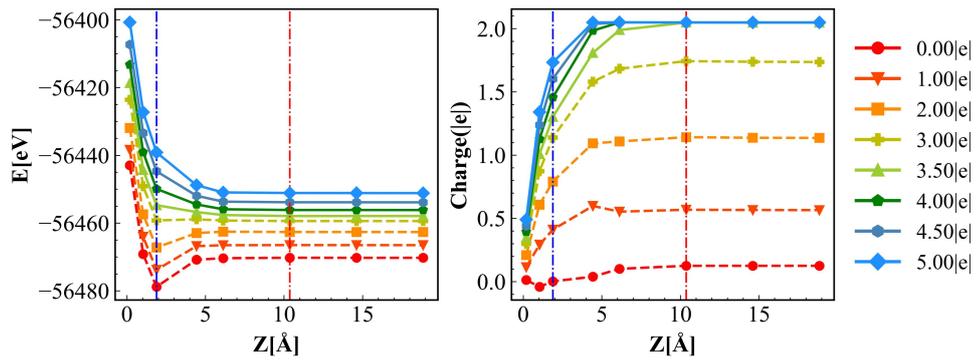

Y(0001):

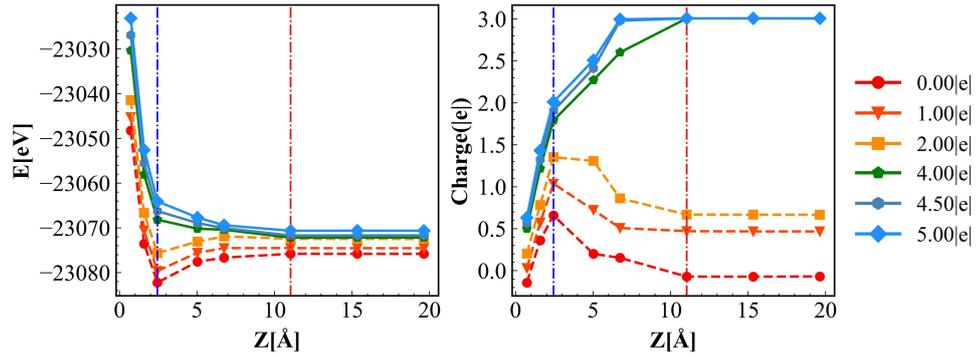

**Category #3: Charge-dependent non-integer valence states after dissolution, challenging the conventional understanding.**

Sc(0001)：

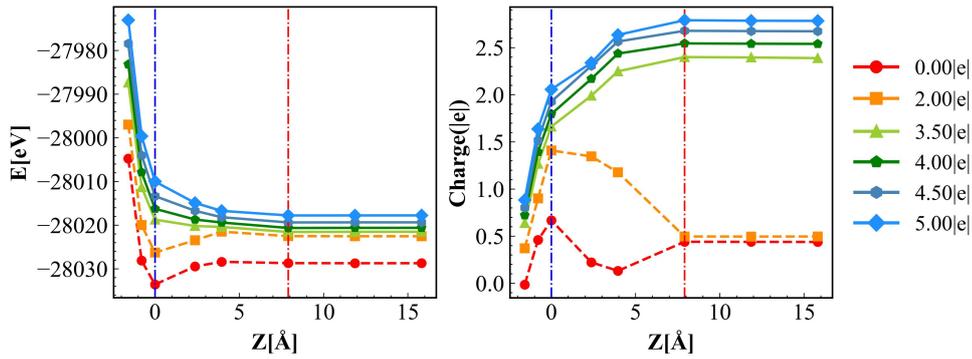

Ag(100)：

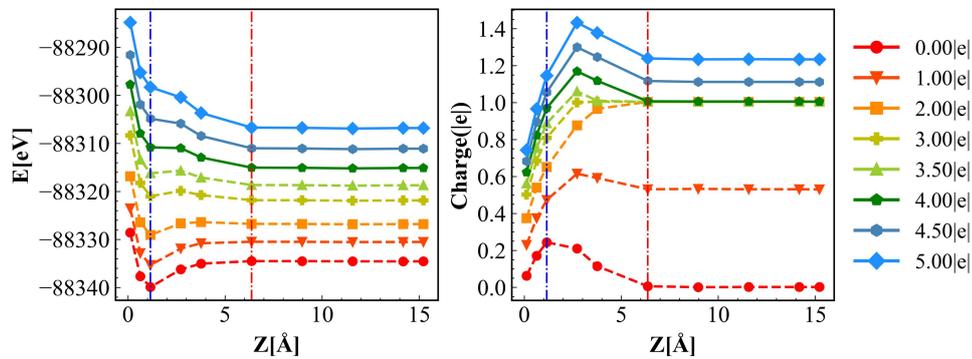

Table1: Comparison of valence state in energy-stable region in systems with $2 \times 2 \times 5$ atoms and $3 \times 3 \times 3$ atoms in the primary computational cell.

| Element | Size(2 × 2 × 5)/\|e\| (Size(3 × 3 × 3)) | | | | | | | | | |
|---|---|---|---|---|---|---|---|---|---|---|
| | 0.00 | 1.00 | 2.00 | 3.00 | 3.50 | 4.00 | 4.50 | 5.00 | 6.00 | 7.00 |
| K | 1.00 | 1.00 | 1.00 | 1.00 | - | 1.00 | - | 1.00 | - | - |
| Li | - | 1.00 | 1.00 | 1.00 | 1.00 | 1.00 | 1.00 | 1.00 | - | - |
| Na | 1.00 | 1.00 | 1.00 | 1.00 | 1.00 | 1.00 | 1.00 | 1.00 | - | - |
| Be | - | - | - | 2.00 | 2.00 | 2.00 | 2.00 | 2.00 | - | - |
| Ca | - | - | - | 2.00 | 2.00 | 2.00 | 2.00 | 2.00 | - | - |
| Mg | 0.00(0.00) | - | - | 2.00(2.00) | 2.00 | 2.00(2.00) | 2.00 | 2.00 | - | - |
| Al | 0.30(1.00) | 1.00(1.00) | 1.00(1.00) | 1.00(1.00) | 1.00 | 1.00(1.00) | - | 3.00 | - | (3.00) |
| Zn | 0.00 | 0.10 | 0.60(0.30) | 1.10 | 1.40 | 1.80(1.20) | 2.00(1.60) | 2.00(2.00) | (2.00) | - |
| Cd | 0.00(0.00) | 0.10 | 0.50(0.10) | 1.00(0.40) | 1.30 | 1.60 | 1.90(1.10) | 2.00(1.40) | (2.00) | - |
| Pb | 0.10 | 0.60 | 1.30 | 2.00 | - | - | 2.00 | 2.00 | - | - |
| Ge | 0.10 | 0.60 | 1.10 | 1.70 | 2.00 | 2.00 | 2.00 | 2.00 | - | - |
| Y | -0.10 | 0.50 | 0.70 | - | - | 3.00 | 3.00 | 3.00 | - | - |
| Cu | 0.30(0.50) | 0.70(1.00) | 1.20(1.20) | 1.30(1.20) | 1.30 | 1.40(1.30) | 1.50 | 1.60(1.40) | (1.50) | - |
| Sc | 0.40 | - | 0.50 | - | 2.40 | 2.50 | 2.70 | 2.80 | - | - |
| Ag | 0.00 | 0.50 | 1.00 | 1.00 | 1.00 | 1.00 | 1.00 | 1.10 | 1.20 | - |